\documentclass[superscriptaddress,aps,preprintnumbers,amsmath,showpacs,amssymb,prd,nofootinbib,reprint]{revtex4-1}

\usepackage[dvipdfmx]{graphicx}
\usepackage{amsfonts}
\usepackage[mathscr]{eucal}
\usepackage{ascmac}
\usepackage{dcolumn}
\usepackage{bm}
\usepackage[colorlinks=true,linkcolor=blue,citecolor=blue]{hyperref}
\usepackage{hyperref}
\usepackage{color}
\begin{document}


\newcommand{\vev}[1]{ \left\langle {#1} \right\rangle }
\newcommand{\bra}[1]{ \langle {#1} | }
\newcommand{\ket}[1]{ | {#1} \rangle }
\newcommand{\eV}{ \ {\rm eV} }
\newcommand{\KeV}{ \ {\rm keV} }
\newcommand{\MeV}{\  {\rm MeV} }
\newcommand{\GeV}{\  {\rm GeV} }
\newcommand{\TeV}{\  {\rm TeV} }
\newcommand{\1}{\mbox{1}\hspace{-0.25em}\mbox{l}}
\newcommand{\Red}[1]{{\color{red} {#1}}}

\newcommand{\lmk}{\left(}  
\newcommand{\rmk}{\right)}
\newcommand{\lkk}{\left[}  
\newcommand{\rkk}{\right]}
\newcommand{\lhk}{\left \{ }  
\newcommand{\rhk}{\right \} }
\newcommand{\del}{\partial}  
\newcommand{\la}{\left\langle} 
\newcommand{\ra}{\right\rangle}
\newcommand{\half}{\frac{1}{2}}

\newcommand{\bea}{\begin{array}}
\newcommand{\eea}{\end{array}}
\newcommand{\beq}{\begin{eqnarray}}
\newcommand{\eeq}{\end{eqnarray}}
\newcommand{\eq}[1]{Eq.~(\ref{#1})}

\newcommand{\dd}{\mathrm{d}}
\newcommand{\Mpl}{M_{\rm Pl}}
\newcommand{\mg}{m_{3/2}}
\newcommand{\abs}[1]{\left\vert {#1} \right\vert}
\newcommand{\mphi}{m_{\phi}}
\newcommand{\Hz}{\ {\rm Hz}}
\newcommand{\for}{\quad \text{for }}
\newcommand{\Min}{\text{Min}}
\newcommand{\Max}{\text{Max}}
\newcommand{\Kahler}{K\"{a}hler }
\newcommand{\cphi}{\varphi}
\newcommand{\Tr}{\text{Tr}}
\newcommand{\diag}{{\rm diag}}

\newcommand{\SUf}{SU(3)_{\rm f}}
\newcommand{\Upq}{U(1)_{\rm PQ}}
\newcommand{\Zpq}{Z^{\rm PQ}_3}
\newcommand{\Cpq}{C_{\rm PQ}}
\newcommand{\ubar}{u^c}
\newcommand{\dbar}{d^c}
\newcommand{\ebar}{e^c}
\newcommand{\nubar}{\nu^c}
\newcommand{\Ndw}{N_{\rm DW}}
\newcommand{\Fpq}{F_{\rm PQ}}
\newcommand{\fpq}{v_{\rm PQ}}
\newcommand{\Br}{{\rm Br}}
\newcommand{\Lag}{\mathcal{L}}
\newcommand{\Lqcd}{\Lambda_{\rm QCD}}

\newcommand{\ji}{j_{\rm inf}} 
\newcommand{\jb}{j_{B-L}} 
\newcommand{\M}{M} 
\newcommand{\im}{{\rm Im} }
\newcommand{\re}{{\rm Re} }

\def\lrf#1#2{ \left(\frac{#1}{#2}\right)}
\def\lrfp#1#2#3{ \left(\frac{#1}{#2} \right)^{#3}}
\def\lrp#1#2{\left( #1 \right)^{#2}}
\def\REF#1{Ref.~\cite{#1}}
\def\SEC#1{Sec.~\ref{#1}}
\def\FIG#1{Fig.~\ref{#1}}
\def\EQ#1{Eq.~(\ref{#1})}
\def\EQS#1{Eqs.~(\ref{#1})}
\def\TEV#1{10^{#1}{\rm\,TeV}}
\def\GEV#1{10^{#1}{\rm\,GeV}}
\def\MEV#1{10^{#1}{\rm\,MeV}}
\def\KEV#1{10^{#1}{\rm\,keV}}
\def\blue#1{\textcolor{blue}{#1}}
\def\red#1{\textcolor{blue}{#1}}

\newcommand{\fa}{f_{a}}
\newcommand{\Uh}{U(1)$_{\rm H}$}
\newcommand{\osc}{_{\rm osc}}

\newcommand{\mav}{\left. m_a^2 \right\vert_{T=0}}
\newcommand{\mat}{m_{a, {\rm QCD}}^2 (T)}
\newcommand{\mam}{m_{a, {\rm M}}^2 }


\preprint{
TU-1062; \\
IPMU18-0079; \\
MIT-CTP/5017
}

\title{
Unified Origin of Axion and Monopole Dark Matter, \\
and Solution to the Domain-wall Problem
}

\author{
Ryosuke Sato
}
\affiliation{Department of Particle Physics and Astrophysics, 
Weizmann Institute of Science, POB 26, Rehovot, Israel}

\author{
Fuminobu Takahashi
}
\affiliation{Department of Physics, Tohoku University, 
Sendai, Miyagi 980-8578, Japan} 
\affiliation{Kavli IPMU (WPI), UTIAS, 
The University of Tokyo, 
Kashiwa, Chiba 277-8583, Japan}
\affiliation{Department of Physics, Massachusetts Institute of Technology, 
Cambridge, MA 02139 USA}

\author{
Masaki Yamada
}
\affiliation{
Institute of Cosmology, Department of Physics and Astronomy, 
Tufts University, Medford, MA 02155, USA
}

\date{\today}

\begin{abstract} 
We propose a scenario where the spontaneous breakdown of the Peccei-Quinn symmetry leads to monopole production. Both the axion and the monopole contribute to dark matter, and the Witten effect on the axion mass is a built-in feature. In the KSVZ-type axion model, seemingly different vacua are actually connected by the hidden gauge symmetry, which makes the axionic string unstable and separate into two Alice strings. The Alice string is attached to a single domain wall due to the QCD instanton effect, solving the domain-wall problem. This is in the same spirit of the Lazarides-Shafi mechanism, although the discrete Peccei-Quinn symmetry is not embedded into the center of the original gauge symmetry. In the DFSZ-type axion model, the domain-wall problem is avoided by the Witten effect. If the Peccei-Quinn symmetry is explicitly broken by a small amount, monopoles acquire a tiny electric charge and become mini-charged dark matter. Interestingly, the quality of the Peccei-Quinn symmetry is closely tied to darkness of dark matter.
\end{abstract}

\maketitle

\section{Introduction
\label{sec:introduction}}

The origin of dark matter (DM) remains a mystery in both cosmology and particle physics. Its longevity may be due to conserved charge, 
light mass, or very weak interactions to the standard model. One example is a hidden monopole whose stability is 
guaranteed by its topological charge~\cite{Murayama:2009nj,Baek:2013dwa,Khoze:2014woa}. 
Another is the QCD axion~\cite{Peccei:1977hh,Peccei:1977ur,Weinberg:1977ma,Wilczek:1977pj} 
or axion-like particles whose mass and couplings are suppressed by the decay constant.  

Monopoles arise at a spontaneously symmetry breaking (SSB) of non-Abelian symmetry,
if the vacuum manifold has a non-trivial second homotopy group. For instance, 
an SU(2)$_H$ gauge symmetry is spontaneously broken down to
U(1)$_H$ in the case of 
a simple 't Hooft-Polyakov monopole~\cite{tHooft:1974kcl,Polyakov:1974ek}, and its cosmological implication
was studied in Refs.~\cite{Baek:2013dwa,Khoze:2014woa}. 

The monopole abundance is determined by the correlation length at the SSB,
which depends on the detailed dynamics of the phase transition. 
If the phase transition is of the first order, monopoles can be produced when the expanding bubbles collide.
In the case with a Coleman-Weinberg potential, a monopole with mass of ${\cal O}(10^{10}) \GeV$ can 
explain the observed DM density for a hidden gauge coupling  $g_H = {\cal O}(0.1)$~\cite{Khoze:2014woa}. 
On the other hand, if the phase transition is of the second order,  the monopole
 mass should be of ${\cal O}(10^2)$\,TeV since its abundance is much larger~\cite{Murayama:2009nj,Baek:2013dwa,Khoze:2014woa}.

The QCD axion is a Nambu-Goldstone boson associated with the spontaneous breakdown 
of a global U(1) Peccei-Quinn (PQ) symmetry. The PQ breaking scale sets the axion decay constant $f_a$,
and they are comparable to each other in a simple set-up. 
The QCD axion and axion-like particles have been extensively studied in the literature (see e.g. 
Refs.~\cite{Kim:2008hd,Wantz:2009it,Ringwald:2012hr,Kawasaki:2013ae} for recent reviews).
The observed neutrino burst duration of SN1987A implies that the axion decay constant 
cannot be smaller than $4 \times 10^{8} \GeV$~\cite{Raffelt:2006cw}.
In the early Universe the axion is produced by the misalignment mechanism~\cite{Preskill:1982cy,Abbott:1982af,Dine:1982ah}.
If the PQ symmetry is already broken during inflation, this leads to an upper bound, $f_a \lesssim 10^{12}$\,GeV, 
barring fine-tuning of the initial displacement.\footnote{
The upper bound is greatly relaxed for low-scale inflation with the Hubble parameter comparable to or less than
the QCD scale~\cite{Graham:2018jyp,Guth:2018hsa}.
}
If the PQ symmetry gets spontaneously broken after inflation, on the other hand, axions are produced by the decay of domain 
walls and cosmic strings. 
In this case the axions explain the observed DM abundance if the PQ breaking scale is 
of ${\cal O}(10^{10}) \GeV$~\cite{Kawasaki:2014sqa} or slightly larger~\cite{Klaer:2017ond}. 

Recently, it was pointed out that there is an interesting interplay between the QCD axion and 
hidden monopoles~\cite{Kawasaki:2015lpf,Nomura:2015xil, Kawasaki:2017xwt};
if the QCD axion is coupled to a hidden U(1)$_H$ under which the hidden monopole is charged, 
the Witten effect induces an effective axion mass in the presence of monopoles~\cite{Witten:1979ey, Fischler:1983sc}. 
The effective axion mass squared is proportional to the monopole
number density which decreases as the Universe expands.
Therefore, the Witten effect is important only in the early Universe, and does not spoil 
the PQ mechanism in the present vacuum. Interestingly, one can solve cosmological problems of the 
QCD axion such as the overabundance/isocurvature problem and the domain wall problem by making use of the
Witten effect~\cite{Kawasaki:2015lpf,Nomura:2015xil, Kawasaki:2017xwt}.

The proximity of the two symmetry breaking scales in the monopole and axion DM scenarios 
implies that they may have a common origin. In this paper, we pursue this possibility and build a simple
model that unifies the origin of the QCD axion and monopole. 
The Witten effect on the axion mass is a built-in feature of this model.
We also find that the domain wall problem of the QCD axion can be avoided 
either by the Witten effect or by the way similar to Lazarides-Shafi mechanism~\cite{Lazarides:1982tw}. 
Interestingly, the high-quality of the PQ symmetry
is closely related to darkness of DM. In other words, if one introduces small PQ symmetry breaking terms,
monopoles acquire a  tiny electric charge and become mini-charged DM, which can be searched for by
e.g. direct DM search experiments. 

The rest of this paper is organized as follows. In the next section we provide a simple model in which the PQ symmetry
breaking generates 't Hooft-Polyakov monopoles in a hidden sector. We study the DM abundances in Sec.~\ref{sec:DM}, and 
observational signatures in Sec.~\ref{sec:signals}. The last section is devoted for discussion and conclusions.

\section{Model
\label{sec:model}}

We introduce a hidden SU(2)$_H$ gauge symmetry and an adjoint complex scalar field $\Phi$. 
We further impose a global U(1)$_{\rm PQ}$ symmetry on $\Phi$, 
which will play a role of the PQ scalar. We assume that U(1)$_{\rm PQ}$ symmetry 
is anomalous under SU(3)$_c$ gauge symmetry for the PQ mechanism to work.

The potential for $\Phi$ is given by 
\begin{align}
 V(\Phi) =& - \frac{M^2}{2} \Tr \,\Phi \Phi^\dagger
 + \frac{\lambda_1}{4} (\Tr\, \Phi \Phi^\dagger)^2 + \frac{\lambda_2}{4} | \Tr\, \Phi^2 |^2, 
 \label{L}
\end{align}
where $\Tr$ represents the trace over the SU(2)$_H$ indices, 
$\lambda_1 + \lambda_2$ is taken to be positive for the potential to be  bounded below, and
we assume $M^2 > 0$ and $\lambda_2 < 0$ to obtain the successful spontaneous breaking of the PQ symmetry
and the monopole solution\footnote{
For $\lambda_2>0$, the SU(2)$_H$ gauge symmetry is completely broken and no monopole solution exists.
In the marginal case, monopoles are created at the first phase transition and 
they survive for some time until the residual U(1)$_H$ gauge symmetry is spontaneously broken by 
the second phase transition.
}.
The vacuum expectation value (VEV) of $\Phi$ is given as
\beq
\label{vac}
 \la \Phi \ra = \frac{v}{\sqrt 2}
 \left(
 \bea{cc}
 1 & 0 \\
 0 & -1
 \eea
 \right),\quad
v = \frac{M}{\sqrt{\lambda_1 + \lambda_2}}.
\eeq
This spontaneously breaks SU(2)$_H$ gauge symmetry down 
to U(1)$_H$.  
At the same time, the U(1)$_{\rm PQ}$ global symmetry is spontaneously broken,
leading to the presence of the QCD axion. The SSB scale $v$ therefore sets both the monopole mass and
the PQ breaking scale (i.e., axion decay constant). 

In addition to the U(1)$_H$ symmetry,
the above VEV of $\Phi$  respects a $Z_2$ symmetry which is generated by $\exp(i\pi T_2)\exp(i\pi Q_{\rm PQ})$,
where $T_2$ is a broken generator of SU(2)$_H$ and $Q_{\rm PQ}$ is the generator of U(1)$_{\rm PQ}$.
This $Z_2$ transformation does not commute with the U(1)$_H$ gauge transformation.
Therefore, the unbroken symmetry in an era between the PQ phase transition and the QCD phase transition 
is U(1)$_H \rtimes Z_2$ symmetry.\footnote{
The same symmetry breaking pattern has been studied in different contexts \cite{Leonhardt:2000km, Chatterjee:2017jsi}.
}
Our model is a variant of the Alice electrodynamics \cite{Kiskis:1978ed, Bucher:1992bd}.

The structure of the vacuum can be parametrized by using a real unit vector $\vec a$ in a 
three-dimensional space and a phase $\phi$ in U(1)$_{\rm PQ}$.
However, $(\vec a, \phi)$ and $(-\vec a, \phi+\pi)$ are identical point because of the $Z_2$ symmetry  described above.
Therefore, the global structure of the vacuum is $(S_2 \times S_1)/Z_2$.
As we expected, this manifold has a non-trivial second homotopy group.
This is the monopole solution associated with the SU(2)$_H \to$ U(1)$_H$ breaking.
We also have two types of cosmic strings. 
One is the usual axionic cosmic string, where the phase of $\Phi$ is rotated by U(1)$_{\rm PQ}$ transformation around the cosmic string. 
The other is a cosmic string where the phase of $\Phi$ is rotated by $\pi$ due to
U(1)$_{\rm PQ}$ transformation and then its sign is changed by the SU(2)$_H$ gauge transformation.
For example, we can consider the following configuration around the string:
\begin{align}
 \la \Phi \ra &=
\frac{v e^{i\theta/2}}{\sqrt 2} \times
 \left(
 \bea{cc}
 \cos(\theta/2) & \sin(\theta/2) \\
 \sin(\theta/2) & -\cos(\theta/2)
 \eea
 \right),
\label{eq:alice_string_configuration}
\end{align}
where $\theta$ is the azimuthal angle of the cylindrical coordinate around the cosmic string.
This is the so-called  Alice string \cite{Schwarz:1982ec, Blinnikov:1982eh}, where the cosmic string consists both of global symmetry and local symmetry. 
As we will see later, this Alice string has a crucial role in the evolution of the string-wall system.

The scalar field $\Phi$ has six degrees of freedom: 
two degrees are eaten by the longitudinal component of the $W_H^\pm$ bosons, 
two form a massive U(1)$_H$ charged scalar $\Phi^\pm$, 
one is the QCD axion $a$, 
and the rest one is a massive Higgs boson $\phi$.  Here the superscripts denote the U(1)$_H$ charge. 
The masses of these heavy gauge  scalar bosons are
\begin{align}
m_{W_H^\pm} & = \sqrt{2} g_H v,\\
m_{\phi}^2 &= 2(\lambda_1 + \lambda_2) v^2, \\
m_{\Phi^\pm}^2 &= 2 |\lambda_2| v^2,
\end{align}
where $g_H$ is the SU(2)$_H$ gauge coupling. 

We will also introduce couplings of $\Phi$ with
the PQ quarks or hidden fermions.
After integrating out them (as well as the standard model (SM) quarks in the DFSZ-type axion model), 
we obtain the axion coupling to gluons and hidden photons at low energy,
\begin{align}
N_{\rm DW}^{(c)}\frac{\alpha_s}{8\pi}\frac{a}{v} G_{a\mu\nu} \tilde{G}^{a\mu \nu} 
+ N_{\rm DW}^{(H)}\frac{\alpha_H}{8\pi}\frac{a}{v} F'_{\mu\nu} \tilde{F}'^{\mu \nu}, 
\label{anomaly}
\end{align}
where $G_{a\mu\nu}$ and $F'_{\mu\nu}$ are the field strengths of gluons and U(1)$_H$ hidden photon, respectively,
their duals are shown with tildes,
$\alpha_s$ and $\alpha_H = g_H^2/4\pi$ denote the strong coupling constant and the fine-structure constant of U(1)$_H$, respectively,
and $N_{\rm DW}^{(c)}$ and $N_{\rm DW}^{(H)}$ are the so-called domain wall numbers specified below. 

The axion decay constant $f_a$ is defined by $f_a = v/N_{\rm DW}^{(c)}$.
The axion coupling to gluons gives rise to the axion potential so that axion is stabilized at the strong CP conserving point, 
solving the strong CP problem.
The coupling to hidden photons usually does not give any mass to the axion, but in the presence of hidden monopoles,
it generates the effective axion mass that depends on the monopole number density via the Witten effect~\cite{Witten:1979ey, Fischler:1983sc}.

\subsection{KSVZ axion model}

Now we shall couple $\Phi$ to colored particles required for the PQ mechanism.
In this subsection, we consider a variant of the KSVZ axion model~\cite{Kim:1979if,Shifman:1979if} where 
heavy quarks are charged under U(1)$_{\rm PQ}$, while the SM quarks are not. 
Matter contents and charge assignments are summarized in Table~\ref{table1}. 
Here we introduce a doublet scalar field $H$ charged under SU(2)$_H$, 
in addition to the bi-fundamental quarks $Q$, $\bar Q$ charged under 
SU(3)$_c \times$SU(2)$_H$, for the reason that will become clear below.

{\renewcommand\arraystretch{1.2}
\begin{table}[t]
\begin{center}
\caption{Charge assignments  in the KSVZ-type model.
\label{table1}}
\begin{tabular}{p{2cm}p{1.5cm}p{1.5cm}p{1.5cm}p{0.75cm}}
\hline
\hline
& $Q$ &$ \bar{Q} $& $\Phi$ & $H$ \\
\hline
SU(3)$_c$ & ${\bm 3}$ & $\bar{\bm 3}$ & ${\bm 1}$ & ${\bm 1}$\\
SU(2)$_H$ & ${\bm 2}$ & $\bar{\bm 2}$ & ${\bm 3}$ & ${\bm 2}$\\
U(1)$_Y$ & $-2/3$ & $2/3$ & $0$ & $0$\\ 
U(1)$_{\rm PQ}$ & $0$ & $-1$ & $1$ & $0$\\
\hline \hline
\end{tabular}\end{center}
\end{table}
}

The renormalizable interactions for $\Phi$, $\bar{Q}$, and $Q$ that are allowed by the symmetries are 
\begin{align}
 \mathcal{L}_{\rm int}
=  y\, \Tr\, \bar{Q} \Phi Q
  + {\rm h.c.} 
\end{align}
For couplings of order unity, the masses of $Q$, $\bar Q$, $W_H^\pm$, $\Phi^\pm$, and 
$\phi$ are of order the SSB scale, $v$.
The axion resides in the phase component of $\Phi_3$ for the vacuum (\ref{vac}). Then, $Q$ and $\bar Q$ play 
the role of the PQ quarks that connect the axion to gluons and hidden photons.
In the present model, the domain wall numbers are given by
$N_{\rm DW}^{(c)} = 2$ and $N_{\rm DW}^{(H)}=3$.  

We assume that the mass of $H$, $m_H$,  is parametrically smaller than the SSB scale $v$.
The primary reason to introduce such light doublet $H = (H^{+1/2}, H^{-1/2})$ is to deplete the abundance 
of $W_H$ and $\Phi$, which
would easily exceed the DM abundance by many orders of magnitude. 
Indeed, the massive $W_H^\pm$ gauge boson and $\Phi^\pm$ scalar boson
can decay into the doublet field after the SSB as
$W_H^+ \to H^{+1/2} (H^{-1/2})^\dagger$ and $\Phi^+ \to a H^{+1/2} (H^{-1/2})^\dagger$.
In the next section we will estimate the mass of $H$ in order to be consistent with the observed 
DM abundance.

The PQ quarks are massless before the SSB, and so they are in thermal equilibrium. 
If the PQ quarks $Q$ and $\bar{Q}$ did not have any interactions with the SM quarks 
other than the gauge interactions, they would be stable and easily overclose the Universe.
In order to make them unstable, 
we introduce an interaction between the PQ quarks and the SM quarks as
\beq
 \mathcal{L}_{\rm int, SM} =
 y_h H Q \bar{d}
  + {\rm h.c.}, 
 \label{L2}
\eeq
where $d$ collectively represents a right-handed down-type quark in the Standard Model, and
we have suppressed the flavor index.  Note that $Q$ and $\bar{Q}$ are charged under U(1)$_Y$ 
so that the interaction term is gauge invariant. In the presence of the above interaction,
$Q$ and $\bar Q$ quickly decay into $H$ and the SM quark after the SSB.

\subsection{DFSZ axion model}

Next, we consider a variant of the DFSZ axion model~\cite{Dine:1981rt,Zhitnitsky:1980tq} 
where the SM quarks are charged under the PQ symmetry.

We extend the SM Higgs sector to a two-Higgs doublet model by introducing
 $h_u$ and $h_d$. 
We also introduce a pair of doublet fermions $\Psi$ and $\bar{\Psi}$ charged under SU(2)$_H$. 
Matter contents and charge assignments are summarized in Table~\ref{table2}.

{\renewcommand\arraystretch{1.2}
\begin{table}[t]
\caption{Charge assignments in the DFSZ-type model.
\label{table2}}
\begin{center}
\begin{tabular}{p{2cm}p{1cm}p{1cm}p{1cm}p{1cm}p{0.75cm}}
\hline
\hline
& $h_u$ &$ h_d$& $\Phi$ & $\Psi$ & $\bar{\Psi}$ \\
\hline
SU(2)$_H$ & ${\bm 1}$ & ${\bm 1}$ & ${\bm 3}$ & ${\bm 2}$ & $\bar{{\bm 2}}$\\
U(1)$_Y$ & $1$ & $-1$ & $0$ & $0$ & $0$\\ 
U(1)$_{\rm PQ}$ & $-1$ & $-1$ & $1$ & $-1$ & $-1$\\
\hline \hline
\end{tabular}\end{center}
\end{table}
}

The interactions for $\Phi$, $h_u$, $h_d$, $\bar{\Psi}$, and $\Psi$ that are allowed by the symmetries are 
\begin{align}
 \mathcal{L}_{\rm int}
=  \lambda_{h2} h_u h_d \Tr \Phi^2 
+ \frac{y_\Psi}{\Mpl} \, \Tr\, \bar{\Psi} \Phi^2 \Psi
  + {\rm h.c.} 
\end{align}
After the PQ symmetry breaking,
the hidden fermions $\Psi$ and $\bar{\Psi}$ obtain 
the mass of order $y_\Psi v^2 / \Mpl$, which is around the electroweak scale for $y_\Psi = {\cal O}(1)$. 
We assume that $\lambda_{h2}$ is sufficiently small so that the interaction term does not affect the successful 
electroweak symmetry breaking by the Higgs doublets.

The SM quarks play 
a role of the PQ quarks that connect the axion to gluons. Similarly, 
integrating out $\Psi$ and $\bar{\Psi}$, the axion is coupled to the hidden photons via the anomaly. 
In the present model, the domain wall numbers are therefore given by
$N_{\rm DW}^{(c)} = 6$ and $N_{\rm DW}^{(H)}= 2$.  

The massive $W_H^\pm$ gauge boson and $\Phi^\pm$ scalar boson can decay into the doublet fermion $\Psi$ 
after the SSB.  Note that there is no heavy PQ quark in this model.

\section{Cosmological domain wall problem
\label{sec:DW}}

Axionic cosmic strings are formed when U(1)$_{\rm PQ}$ is spontaneously broken at the SSB.
If there were not for the SU(2)$_H$ gauge symmetry and the monopoles,
the domain wall number $N_{\rm DW}^{(c)} \ne 1$ would imply that there are $N_{\rm DW}^{(c)}$
 degenerate vacua below the QCD scale. After the QCD phase transition, 
each axionic cosmic string would be attached to $N_{\rm DW}^{(c)}$ domain walls. 
Such string-wall network is stable and their energy density comes to dominate the Universe at the end of the day, making the Universe
intolerably inhomogeneous. However, the domain-wall number and the string-wall dynamics
are drastically changed if we correctly take into account the SU(2)$_H$ gauge symmetry and the presence of monopoles in the Universe. 
In this section we explain how they solve the cosmological domain-wall problem in the two axion models given in the previous section.

\subsection{Lazarides-Shafi mechanism}

Suppose that all degenerate QCD vacua are connected by a local or global symmetry $A$, 
which is spontaneously broken at the same time with (or below the energy scale of) the PQ symmetry. 
In this case, the energetically most favorable configuration of cosmic string 
is the one around which 
the PQ phase (i.e., the axion field value divided by $f_a$) changes by a factor of $2 \pi / N_{\rm DW}^{(c)}$ 
and the rest of the phase for $\Phi$ is complemented by the NG boson of the symmetry $A$. 
Such a cosmic string will be attached by only one domain wall due to the QCD instanton effect. 
In this case the domain-wall number is effectively reduced to be unity. 
Then those topological defects tend to shrink to a point because of the tension of domain walls and are unstable. 
This is essentially the mechanism proposed by Lazarides and Shafi for the solution to the domain wall problem~\cite{Lazarides:1982tw} (see Ref.~\cite{Kawasaki:2015ofa} for a recent work).

Let us emphasize here that the discrete symmetry that transforms the degenerate vacua into themselves 
does not have to be embedded into the center of the group $A$. 
The necessary and sufficient condition for the mechanism to work 
is that the degenerate vacua are connected by another symmetry than U(1)$_{\rm PQ}$. 
In any case we call the above solution the Lazarides-Shafi mechanism in this paper.

In the case of the KSVZ-type axion model, the domain wall number is $N_{\rm DW}^{(c)} = 2$. 
The degenerate vacua are transformed into each other via the sign flip: $\Phi \to - \Phi$. 
This transformation can be realized by 
the SU(2)$_H$ gauge transformation 
as well as the U(1)$_{\rm PQ}$ transformation. 
Hence the SU(2)$_H$ symmetry plays the role of what we denote as $A$ above 
and 
the degenerated vacua are connected with each other via the (spontaneously broken) SU(2)$_H$ gauge transformation.

As we have seen in the previous section, we have two types cosmic string; the axionic string and the Alice string.
While the axionic string will be attached by two domain walls by the QCD instanton effect,
the Alice string will be attached by one domain wall
because the U(1)$_{\rm PQ}$ phase rotation around the string is only $\pi$ (see Eq.~(\ref{eq:alice_string_configuration})).
In particular, the usual axionic string is unstable and separate into a pair of the Alice strings. 
The cosmological domain-wall problem is therefore avoided by the Lazarides-Shafi mechanism. 

In the DFSZ-type axion model, on the other hand, the domain wall number is 
$N_{\rm DW}^{(c)} = 6$, and there are still three physically distinct vacua after
taking account of the SU(2)$_H$ gauge symmetry.
In this case, one needs another mechanism to avoid the cosmological domain-wall problem,
which will be discussed next.

\subsection{Witten effect}

Here we focus on the DFSZ-type axion model and the cosmological domain-wall problem
is intrinsic to this model. Below we will see the domain-wall problem 
can be solved by the Witten effect. 

As shown in the next section, 
monopoles form at the SSB of the SU(2)$_H$ symmetry. 
The axion acquires an effective mass due to the Witten effect in the 
presence of monopoles even before the QCD phase transition~\cite{Kawasaki:2015lpf,Nomura:2015xil, Kawasaki:2017xwt}. 
In particular, the U(1)$_{\rm PQ}$ symmetry is explicitly broken by the Witten effect 
down to $Z_2$ discrete symmetry because of  $N_{\rm DW}^{(H)} = 2$.
These degenerate vacua are connected by the SU(2)$_H$ symmetry 
and therefore,  the effective domain-wall number due to the Witten effect is reduced to unity in combination with
 the Lazarides-Shafi mechanism.  As a result, 
each cosmic string (of the Alice type) is attached by one domain wall. 
The tension of those domain walls are determined by the mass of axion 
as 
\beq
 \sigma_a (t) \sim c_a m_{a,M} (t) f_a^2, 
\eeq
where $c_a = {\cal O}(1)$ is a numerical constant, and $m_{a,M} (t)$ is the axion mass due to the Witten effect.\footnote{
The effective axion mass depends on the spatial distribution of the monopoles
around the cosmic string. Also, the axion potential has a spinodal point at the maximum, and
we make only an order of magnitude estimate here. 
}

The effective axion mass $m_{a,M} (t)$ is determined by the number density of monopoles as~\cite{Fischler:1983sc}
\beq
 m_{a,M}^2 \simeq 2 \kappa \frac{n_M (T)}{f_a}, 
\eeq
where 
\beq
 \kappa \equiv \lmk \frac{N_{\rm DW}^{(H)}}{N_{\rm DW}^{(c)} } \rmk^2 
 \frac{\alpha_H }{ 32 \pi^2 r_c f_a}.
\eeq
Here 
$r_c = m_\Psi^{-1}$ is the electric screening scale due to $\Psi$, 
and $\kappa$ is of order $10^{-10}$ for $f_a \simeq 10^{10}$\,GeV, $m_\Psi \simeq 1$\,TeV
and $\alpha_H \simeq 0.1$.
Domain walls form when 
the axion mass becomes larger than the Hubble parameter. 
This happens at the temperature of 
\beq
 T_{DW} &\simeq& 
 0.02 \GeV \lmk \frac{\kappa}{10^{-10}} \rmk
 \lmk \frac{m_M}{10^{10} \GeV} \rmk^{-1} 
 \nonumber\\
 &&~~~~\times \lmk \frac{\Omega_M h^2}{0.12} \rmk 
 \lmk \frac{f_a}{10^{10} \GeV} \rmk^{-1}. 
 \label{T_DW}
\eeq
This is around the QCD scale, so that this effective mass due to the Witten effect 
can be neglected until the QCD phase transition occurs.

After the QCD phase transition, 
the axion also acquires a mass from the nonperturbative QCD effects.
At zero temperature, it is given by 
\beq
 m_a \simeq \frac{\sqrt{m_u m_d}}{m_u + m_d} \frac{m_\pi f_\pi}{f_a}, 
\eeq
where $m_u$, $m_d$, and $m_\pi$ are respectively up quark, down quark, and pion masses, 
and $f_\pi$ is the pion decay constant. 
The domain wall number of the axion coupling to gluons is given by $N_{\rm DW}^{(c)} = 6$, 
so that this non-perturbative effect breaks U(1)$_{\rm PQ}$ down to $Z_6$ discrete symmetry. 
The degenerate six vacua get into three pairs by the SU(2)$_H$ symmetry 
and the effective domain-wall number is reduced to three by the Lazarides-Shafi mechanism. 

Since the two effective domain-wall numbers of the axion couplings to U(1)$_H$ and SU(3)$_c$ are 
relatively prime ($N_{\rm DW}^{(H)}/2 =1$ and $N_{\rm DW}^{(c)}/2 = 3$), the PQ symmetry is completely broken by the Witten effect and the nonperturbative QCD effect.
Thus, the vacuum degeneracy is lifted in the presence of monopoles at a scale below the QCD scale. 
As a result, domain walls experience a pressure from the vacuum energy in the false vacua and 
shrink to a point and the entire space will be filled with the true vacuum (which approaches to the strong CP conserving
minimum as the Witten effect becomes inefficient).
They disappear when 
\beq
 \sigma_a (t) = \frac{\mu(t)}{t}, 
\eeq
or $T \sim T_{DW}$, 
where 
$\mu(t) \simeq \pi f_a^2 \ln \lmk \delta_s^{-1} t \rmk$ is 
the tension of the cosmic string 
and $\delta_s$ is its core width. 
Since $T_{DW}$ is about the QCD scale, 
the domain wall and cosmic string system disappear soon after the QCD phase transition.

\section{Dark matter candidates
\label{sec:DM}}

In the setup described in Sec.~\ref{sec:model}, the monopole is the only stable object whose mass is 
of order or higher than the SSB scale. Below the SSB scale, we have the following stable particles: 
U(1)$_H$ charged fields ($H^{\pm 1/2}$ in the KSVZ-type model and $\Psi^{\pm 1/2}$ in the DFSZ-type model), the QCD axion $a$, and U(1)$_H$ gauge field in order of
decreasing mass. While the U(1)$_H$ gauge boson contributes to dark radiation,
the other particles are massive and contribute to DM. In the following we estimate the contributions
to the DM abundance.

\subsection{Hidden monopoles}

We consider the case where the phase transition is of the first order. In this case the
phase transition ends by the bubble coalescence. 
The monopole abundance is determined by the Kibble mechanism~\cite{Kibble:1976sj}:
the orientation of the scalar field is random at scales beyond the correlation length, and so,
monopoles are created with a probability close to unity when the bubbles collide. 

The bubble nucleation rate per unit volume per unit time 
is given by $\Gamma_{\rm bubble} \sim T^4 e^{-S_3/T}$ at a finite temperature $T$, 
where $S_3$ is the three-dimensional Euclidean action of a bubble solution.
The phase transition starts when the bubble nucleation rate 
in a Hubble volume 
becomes comparable to the Hubble expansion rate $H(t)$: 
\beq
 H^4(t_c) \sim T_c^4 e^{-S_3/T_c}. 
 \label{t_c}
\eeq
where $t_c$ and $T_c=T(t_c) $ are the time and temperature at the beginning of the bubble
nucleation, respectively. 

After the nucleation, the bubble wall expands at a velocity close to the speed of light. 
The phase transition completes 
when the entire space is filled with the bubbles. 
This time scale is given by the inverse of the time derivative 
of the nucleation rate: 
\beq
 \frac{d}{d t} \ln \lmk  T_c^4 e^{-S_3/T_c} \rmk_{t = t_c} 
 \sim H_c \beta, 
\eeq
where we define 
\beq
 \beta \equiv - \left.\frac{1}{H} \frac{d}{d t} \frac{S_3}{T} \right|_{t=t_c},
\eeq
and we assume $\beta \gg 1$ in the second equality. 
In fact, from \eq{t_c}, 
we expect 
\beq
 \beta \sim \frac{S_3}{T} (t_c) \simeq \ln \lmk \frac{T_c^4}{H_c^4} \rmk,
\eeq
where $H_c \sim T_c^2 / \Mpl$, and so, for $T_c \ll \Mpl$, 
we find $\beta \gg 1$. 

Since the phase transition completes 
within the time scale of order $(H_c \beta)^{-1}$ after $t=t_c$,
the typical size of the bubbles at the end of phase transition is about $(H_c \beta)^{-1}$. 
Hence the number of nucleated bubbles within a Hubble volume is given by 
$\sim \beta^3$. 
Since  monopoles are created when the bubbles collide,
the monopole number density after the phase transition is given by 
\beq
 \frac{n_M}{s} \sim \frac{\beta^3 H_c^3}{T_c^3}. 
\eeq
Thus, the monopole abundance is given by 
\beq
 \Omega_M h^2 \sim 0.1
 \lmk \frac{m_M}{10^{10} \GeV} \rmk 
 \lmk \frac{T_c}{3 \times 10^9 \GeV} \rmk^3 
 \lmk \frac{\beta}{80} \rmk^3, 
 \nonumber\\
\eeq
where $m_M$ is the monopole mass. 

To solve the domain-wall problem in the DFSZ-type axion model, 
the monopole abundance should be sufficiently large (see \eq{T_DW}). 
This put a lower bound on the scale of the symmetry breaking. 
The domain walls should disappear before the Big-Bang nucleosynthesis starts. 
Hence we require $T_{\rm DW} \gtrsim {\cal O}(1) \MeV$, 
which gives $v \, (\propto T_c) \gtrsim 10^9 \GeV$.

One may consider the Coleman-Weinberg type potential 
to realize the strong first order phase transition. 
In this case, the breaking scale $v$ is larger than the critical temperature $T_c$ 
by a factor about $1/g_H$. Hence we expect that the breaking scale is of order $10^{10} \GeV$, 
which is also reasonable for the PQ breaking scale (cf. the next subsection).

\subsection{Non-relativistic axions}

As we have seen in Sec.~\ref{sec:DW}, 
the domain walls and cosmic strings disappear at the QCD phase transition in our models. 
During the annihilation process of domain walls and cosmic strings, 
their energy will be released as marginally relativistic axions. 
Those axions will be non-relativistic soon after the emission. 
The resulting energy density of axions 
can be estimated by~\cite{Kawasaki:2014sqa}
\beq
 \Omega_a h^2 \sim 10^{-2} \lmk \frac{f_a}{10^{10} \GeV} \rmk^{1.19}. 
\eeq
This is consistent with the experimental lower bound on the axion decay constant~\cite{Raffelt:2006cw},
and the PQ breaking scale happens to be close to the monopole mass to explain DM.

\subsection{Charged hidden field}

The abundance of U(1)$_H$ charged fields ($H^{\pm 1/2}$ in the KSVZ-type model 
and $\Psi^{\pm 1/2}$ in the DFSZ-type model) is determined by the freeze-out mechanism. 
Here we focus on the abundance of $H^{\pm 1/2}$, and the result is similar for $\Psi$
(except for the argument about the Higgs portal coupling). 

The abundance of $H$ is given by 
\beq
 \Omega_H h^2 \approx \frac{5.0 \times 10^{-27} {\rm \ cm}^3 {\rm s}^{-1}}{\la \sigma_H v \ra}, 
\eeq
where the annihilation cross section is given by 
\beq
 \la \sigma_H v \ra \simeq \frac{\pi \alpha_h^2 }{m_H^2}. 
 \label{annihilation}
\eeq
Hence we obtain 
\beq
 \Omega_H h^2 \approx 0.01 \lmk \frac{m_H}{1 \ {\rm TeV}} \rmk^2 \lmk \frac{\alpha_H}{0.1} \rmk^{-2}. 
 \label{Omega_L}
\eeq

In the KSVZ-type axion model, 
it is possible to include a Higgs portal coupling,  $\lambda_H |h|^2  \abs{H}^2$, where $h$ denotes
the SM Higgs field.
The direct DM search experiments  put a very tight constraint on 
this coupling and excluded most of the parameter space~\cite{Cline:2013gha} 
unless the annihilation via the U(1)$_H$ gauge interaction is efficient. 
Hence we assume that the Higgs portal coupling is so small that it does not
contribute to the annihilation process.

\section{Observational implications
\label{sec:signals}}

\subsection{Kinetic mixing}

If the PQ symmetry is classically exact, 
the kinetic mixing between U(1)$_Y$ gauge boson 
and U(1)$_H$ gauge boson is absent because of charge conjugation symmetry 
in the U(1)$_H$ sector. 
In the language of high-energy theory, 
the kinetic mixing will be absent because 
the following operator is prohibited by the PQ symmetry: 
\beq
\frac{1}{\Mpl} \Tr \Phi F_Y F_H, 
 \label{mixing}
\eeq
which would give a kinetic mixing of order $v/\Mpl$ after the SSB.

If the PQ symmetry is broken by a higher-dimensional operators, on the other hand,
there appears a small kinetic mixing between U(1)$_h$ and U(1)$_Y$. 
In the rest of this subsection we focus on the KSVZ-type axion model. 
Suppose that 
we have the following higher-dimensional operators:
\beq
 \frac{\lambda}{\Mpl^{n}} \Tr \bar{Q} \Phi^{n+1} Q + {\rm h.c.}, 
\eeq
where $n$ is an integer and $\lambda$ is a coupling constant of order unity. 
This term breaks the PQ symmetry and gives an explicit mass to the axion by quantum corrections.
The axion mass induced by this operator can be estimated as 
\beq 
 m_{a,\lambda}^2 \sim \frac{\lambda}{16 \pi^2} \lmk \frac{v}{\Mpl} \rmk^{n-2} v^2. 
\eeq
To solve the strong CP problem, 
this must be smaller than about $10^{-10} m_a^2$. 
Noting that $v \sim 10^{10} \GeV$, we find that $n$ should be larger than or equal to eight 
to explain the smallness of the strong CP phase.

The same operator violates the mass degeneracy between $Q^1$ and $Q^2$: $\delta m \equiv m_{Q^1} - m_{Q^2} = \lambda v (v/\Mpl)^{n+1}$. 
In this case, the one-loop diagrams associated with these fields lead to a kinetic mixing $\chi$ between U(1)$_Y$ and U(1)$_H$: 
\beq
 \chi \sim \frac{g_H g_Y}{16 \pi^2} \ln \lmk \frac{\delta m_Q}{m_Q} \rmk^2, 
\eeq
where $m_Q = y v$. 
The logarithmic factor gives a factor of $10^{-2}$ or smaller.

At low temperature, the axion VEV cancels the strong CP phase 
but does not cancel the CP phase associated with U(1)$_H$ in the hidden sector (see \eq{anomaly}). 
Hence the monopole generically acquires a hidden electric charge of order unity 
and becomes a dyon in terms of the U(1)$_H$ gauge symmetry. 
As a result, 
the kinetic mixing induces a mini electric charge for the monopole~\cite{Bruemmer:2009ky}. 
The constraint of the kinetic mixing due to the absence of mini-charged particles 
is given by 
\beq
 \chi g_H \lesssim 10^{-6}, 
\eeq
for the monopole DM with mass of $10^{10} \GeV$~\cite{DelNobile:2015bqo, Akerib:2016vxi}. 
This is satisfied when $g_H \lesssim 0.1$ for $n = 3$.

The charged scalar field $H^{\pm 1/2}$ also acquires 
the electric charge via the kinetic mixing. 
If its amount is as large as the observed DM abundance, 
the null-result of mini-charged particle search puts a stronger constraint 
by a factor of $10^{-4}$. 
Since the kinetic mixing depends on the PQ breaking parameter 
only logarithmically, this constraint excludes the scenario with even a tiny amount of the explicit PQ breaking effect. 
Therefore, 
the darkness of DM 
is closely tied to the quality of PQ symmetry in this case.
The high quality of PQ symmetry may be explained by the anthropic argument 
because the large scale structure does not form in the Universe in a scenario of charged DM.

In the case of the DFSZ-type axion model, 
there is no particles that are charged under both U(1)$_H$ and U(1)$_Y$. 
Then the kinetic mixing comes only from the higher-dimensional operator of \eq{mixing}.

\subsection{Dark radiation}

The axion is thermalized after the PQ phase transition~\cite{Turner:1986tb, Masso:2002np}.
The remaining hidden gauge field U(1)$_H$ 
as well as the axion thus contribute to the energy density of the Universe 
as dark radiation. 
The amount of dark radiation $\rho_{\rm DR}$ is conveniently described by the effective neutrino number 
$N_{\rm eff}$ as 
\beq
 \Delta N_{\rm eff} = \frac{4}{7} \frac{\rho_{\rm DR}}{(\pi^2 / 30) T_\nu^4}, 
 \label{delta N_eff}
\eeq
where $T_\nu$ is the neutrino temperature.

First we consider the KSVZ-type  axion model. 
In the absence of the Higgs portal interaction, 
the hidden gauge field U(1)$_H$ and $H^{\pm 1/2}$ are thermalized and are decoupled from the 
SM sector well before the electroweak phase transition. 
Then the latter field annihilates into the former one during the freeze-out epoch. 
The resulting energy density of hidden gauge field 
is thus larger than that of axion by a factor of $2 \times 3^{4/3}$. 
Hence we obtain 
\beq
  \Delta N_{\rm eff} 
  &=& \frac{4}{7} \lmk 2 \times 3^{4/3} + 1 \rmk 
  \lmk \frac{g_* (T_D)}{43/4} \rmk^{-4/3}
  \\
  &\simeq& 0.26, 
 \label{delta N_eff 2}
\eeq
where $g_*$ ($= 106.75$) is the effective relativistic degrees of freedom in the standard model sector 
at the decoupling temperature $T_D$.

If the Higgs-portal interaction is strong enough, 
the hidden sector is in thermal equilibrium around the electroweak scale 
until the density of the charged scalar $H^{\pm 1/2}$ freezes out. 
In this case, the decoupling temperature is about $m_H / 10$ 
and $g_*$ is about $90-100$ for $m_H \gtrsim 100 \GeV$. 
Since the annihilation into the U(1)$_H$ gauge boson dominates, 
the energy density of hidden gauge field 
is again larger than that of axion by a factor of $2 \times 3^{4/3}$ except for the difference of $g_*$. 
Hence we obtain 
\beq
  \Delta N_{\rm eff} 
  &=& \frac{4}{7}
  \lmk \frac{106.75}{43/4} \rmk^{-4/3}
  +\frac{4}{7} 2 \times 3^{4/3} 
  \lmk \frac{90-100}{43/4} \rmk^{-4/3}
  \nonumber\\
  &\simeq& 0.28-0.32. 
 \label{delta N_eff 3}
\eeq

The Planck data combined with the observation of BAO 
puts the constraint $N_{\rm eff} = 3.15 \pm 0.23$~\cite{Ade:2015xua}. 
The standard model prediction is $N_{\rm eff} = 3.046$ 
and hence our result is consistent with the constraint. 
It is expected that the ground-based Stage-IV CMB polarization experiment 
(CMB-S4) will measure $N_{\rm eff}$ with a precision of $\Delta N_{\rm eff} = 0.0156$ 
within $1 \sigma$ level~\cite{Wu:2014hta} (see also Ref.~\cite{Abazajian:2013oma}). 

Next, we consider the DFSZ-type axion model, 
where the hidden gauge field and $\Psi^{\pm 1/2}$ are decoupled from the SM sector well before the electroweak phase transition. 
The effective relativistic degrees of freedom of the charged fermions is given by $2 \times 4 \times 7/8$ at a high temperature. 
After the freeze-out epoch, the effective number of neutrino is given by 
\beq
  \Delta N_{\rm eff} 
  &=& \frac{4}{7} \lmk 2 \times (9/2)^{4/3} + 1 \rmk 
  \lmk \frac{g_* (T_D)}{43/4} \rmk^{-4/3}
  \\
  &\simeq& 0.42, 
\eeq
where we used $g_* = 106.75$. 
This is consistent with the Planck constraint within $2 \sigma$ level.

\section{Discussion and Conclusions
\label{sec:conclusions}}
Motivated by the coincidence of the energy scales,  we have pursued a possibility of unifying the PQ symmetry breaking and the production of the monopole DM.  We have provided both KSVZ- and DFSZ-type axion models, where the cosmological domain-wall problem can be avoided 
either by a mechanism in the same spirit of the Lazarides-Shafi mechanism or by the Witten effect. 

An SU(2) doublet field needs to be introduced to make some unwanted heavy relics unstable. As a result, there are three candidates of DM: 
the axion, monopole, and hidden-charged field. 
The latter two are charged under the remaining U(1) gauge symmetry, 
so that they may acquire a nonzero electric charge via a possible kinetic mixing  between the electroweak and hidden U(1) gauge bosons. 
We have found that the amount of kinetic mixing, or the electric charges of those DM, 
is related to the quality of the PQ symmetry. 
Hence the darkness of DM, which is required by the large scale structure formation, may explain the high quality of the PQ symmetry.

The hidden gauge bosons, as well as the relativistic components of axions, contribute to the energy density of the Universe as dark radiation. 
The amount of those energy density is consistent with the present observational result 
and can be distinguished from the standard scenario in the near future.

Finally, we comment on 
the detectability of gravitational waves that are emitted from the dynamics of bubble at the first-order phase transition. 
Since the energy scale of phase transition is as high as $10^{10} \GeV$, 
the typical frequency of those gravitational waves is too high to be detected by the proposed detectors~\cite{Huang:2018fum}.

%
\section*{Acknowledgments}
F.T. thanks the hospitality of MIT Center for Theoretical Physics where part of this work was done.
R.S. thanks the hospitality of CERN theory group. 
This work is supported by JSPS KAKENHI Grant Numbers JP15H05889 (F.T.), 
JP15K21733 (F.T.), JP17H02878 (F.T.), and JP17H02875 (F.T.), Leading Young Researcher Overseas
Visit Program at Tohoku University (F.T.) and by World Premier International Research Center Initiative (WPI Initiative), 
MEXT, Japan (F.T.). 
%

\vspace{1cm}

\bibliography{reference}

\providecommand{\href}[2]{#2}\begingroup\raggedright\begin{thebibliography}{10}

\bibitem{Murayama:2009nj}
H.~Murayama and J.~Shu, \emph{{Topological Dark Matter}},
  \href{https://doi.org/10.1016/j.physletb.2010.02.037}{\emph{Phys. Lett.}
  {\bfseries B686} (2010) 162}
  [\href{https://arxiv.org/abs/0905.1720}{{\ttfamily 0905.1720}}].

\bibitem{Baek:2013dwa}
S.~Baek, P.~Ko and W.-I. Park, \emph{{Hidden sector monopole, vector dark
  matter and dark radiation with Higgs portal}},
  \href{https://doi.org/10.1088/1475-7516/2014/10/067}{\emph{JCAP} {\bfseries
  1410} (2014) 067} [\href{https://arxiv.org/abs/1311.1035}{{\ttfamily
  1311.1035}}].

\bibitem{Khoze:2014woa}
V.~V. Khoze and G.~Ro, \emph{{Dark matter monopoles, vectors and photons}},
  \href{https://doi.org/10.1007/JHEP10(2014)061}{\emph{JHEP} {\bfseries 10}
  (2014) 061} [\href{https://arxiv.org/abs/1406.2291}{{\ttfamily 1406.2291}}].

\bibitem{Peccei:1977hh}
R.~D. Peccei and H.~R. Quinn, \emph{{CP Conservation in the Presence of
  Instantons}}, \href{https://doi.org/10.1103/PhysRevLett.38.1440}{\emph{Phys.
  Rev. Lett.} {\bfseries 38} (1977) 1440}.

\bibitem{Peccei:1977ur}
R.~D. Peccei and H.~R. Quinn, \emph{{Constraints Imposed by CP Conservation in
  the Presence of Instantons}},
  \href{https://doi.org/10.1103/PhysRevD.16.1791}{\emph{Phys. Rev.} {\bfseries
  D16} (1977) 1791}.

\bibitem{Weinberg:1977ma}
S.~Weinberg, \emph{{A New Light Boson?}},
  \href{https://doi.org/10.1103/PhysRevLett.40.223}{\emph{Phys. Rev. Lett.}
  {\bfseries 40} (1978) 223}.

\bibitem{Wilczek:1977pj}
F.~Wilczek, \emph{{Problem of Strong p and t Invariance in the Presence of
  Instantons}}, \href{https://doi.org/10.1103/PhysRevLett.40.279}{\emph{Phys.
  Rev. Lett.} {\bfseries 40} (1978) 279}.

\bibitem{tHooft:1974kcl}
G.~'t~Hooft, \emph{{Magnetic Monopoles in Unified Gauge Theories}},
  \href{https://doi.org/10.1016/0550-3213(74)90486-6}{\emph{Nucl. Phys.}
  {\bfseries B79} (1974) 276}.

\bibitem{Polyakov:1974ek}
A.~M. Polyakov, \emph{{Particle Spectrum in the Quantum Field Theory}},
  {\emph{JETP Lett.} {\bfseries 20} (1974) 194}.

\bibitem{Kim:2008hd}
J.~E. Kim and G.~Carosi, \emph{{Axions and the Strong CP Problem}},
  \href{https://doi.org/10.1103/RevModPhys.82.557}{\emph{Rev. Mod. Phys.}
  {\bfseries 82} (2010) 557} [\href{https://arxiv.org/abs/0807.3125}{{\ttfamily
  0807.3125}}].

\bibitem{Wantz:2009it}
O.~Wantz and E.~P.~S. Shellard, \emph{{Axion Cosmology Revisited}},
  \href{https://doi.org/10.1103/PhysRevD.82.123508}{\emph{Phys. Rev.}
  {\bfseries D82} (2010) 123508}
  [\href{https://arxiv.org/abs/0910.1066}{{\ttfamily 0910.1066}}].

\bibitem{Ringwald:2012hr}
A.~Ringwald, \emph{{Exploring the Role of Axions and Other WISPs in the Dark
  Universe}}, \href{https://doi.org/10.1016/j.dark.2012.10.008}{\emph{Phys.
  Dark Univ.} {\bfseries 1} (2012) 116}
  [\href{https://arxiv.org/abs/1210.5081}{{\ttfamily 1210.5081}}].

\bibitem{Kawasaki:2013ae}
M.~Kawasaki and K.~Nakayama, \emph{{Axions: Theory and Cosmological Role}},
  \href{https://doi.org/10.1146/annurev-nucl-102212-170536}{\emph{Ann. Rev.
  Nucl. Part. Sci.} {\bfseries 63} (2013) 69}
  [\href{https://arxiv.org/abs/1301.1123}{{\ttfamily 1301.1123}}].

\bibitem{Raffelt:2006cw}
G.~G. Raffelt, \emph{{Astrophysical axion bounds}},
  \href{https://doi.org/10.1007/978-3-540-73518-2_3}{\emph{Lect. Notes Phys.}
  {\bfseries 741} (2008) 51}
  [\href{https://arxiv.org/abs/hep-ph/0611350}{{\ttfamily hep-ph/0611350}}].

\bibitem{Preskill:1982cy}
J.~Preskill, M.~B. Wise and F.~Wilczek, \emph{{Cosmology of the Invisible
  Axion}}, \href{https://doi.org/10.1016/0370-2693(83)90637-8}{\emph{Phys.
  Lett.} {\bfseries B120} (1983) 127}.

\bibitem{Abbott:1982af}
L.~F. Abbott and P.~Sikivie, \emph{{A Cosmological Bound on the Invisible
  Axion}}, \href{https://doi.org/10.1016/0370-2693(83)90638-X}{\emph{Phys.
  Lett.} {\bfseries B120} (1983) 133}.

\bibitem{Dine:1982ah}
M.~Dine and W.~Fischler, \emph{{The Not So Harmless Axion}},
  \href{https://doi.org/10.1016/0370-2693(83)90639-1}{\emph{Phys. Lett.}
  {\bfseries B120} (1983) 137}.

\bibitem{Graham:2018jyp}
P.~W. Graham and A.~Scherlis, \emph{{The Stochastic Axion Scenario}},
  \href{https://arxiv.org/abs/1805.07362}{{\ttfamily 1805.07362}}.

\bibitem{Guth:2018hsa}
A.~H. Guth, F.~Takahashi and W.~Yin, \emph{{The QCD Axion Window and Low Scale
  Inflation}},  \href{https://arxiv.org/abs/1805.08763}{{\ttfamily
  1805.08763}}.

\bibitem{Kawasaki:2014sqa}
M.~Kawasaki, K.~Saikawa and T.~Sekiguchi, \emph{{Axion dark matter from
  topological defects}},
  \href{https://doi.org/10.1103/PhysRevD.91.065014}{\emph{Phys. Rev.}
  {\bfseries D91} (2015) 065014}
  [\href{https://arxiv.org/abs/1412.0789}{{\ttfamily 1412.0789}}].

\bibitem{Klaer:2017ond}
V.~B. Klaer and G.~D. Moore, \emph{{The dark-matter axion mass}},
  \href{https://doi.org/10.1088/1475-7516/2017/11/049}{\emph{JCAP} {\bfseries
  1711} (2017) 049} [\href{https://arxiv.org/abs/1708.07521}{{\ttfamily
  1708.07521}}].

\bibitem{Kawasaki:2015lpf}
M.~Kawasaki, F.~Takahashi and M.~Yamada, \emph{{Suppressing the QCD Axion
  Abundance by Hidden Monopoles}},
  \href{https://doi.org/10.1016/j.physletb.2015.12.075}{\emph{Phys. Lett.}
  {\bfseries B753} (2016) 677}
  [\href{https://arxiv.org/abs/1511.05030}{{\ttfamily 1511.05030}}].

\bibitem{Nomura:2015xil}
Y.~Nomura, S.~Rajendran and F.~Sanches, \emph{{Axion Isocurvature and Magnetic
  Monopoles}},
  \href{https://doi.org/10.1103/PhysRevLett.116.141803}{\emph{Phys. Rev. Lett.}
  {\bfseries 116} (2016) 141803}
  [\href{https://arxiv.org/abs/1511.06347}{{\ttfamily 1511.06347}}].

\bibitem{Kawasaki:2017xwt}
M.~Kawasaki, F.~Takahashi and M.~Yamada, \emph{{Adiabatic suppression of the
  axion abundance and isocurvature due to coupling to hidden monopoles}},
  \href{https://doi.org/10.1007/JHEP01(2018)053}{\emph{JHEP} {\bfseries 01}
  (2018) 053} [\href{https://arxiv.org/abs/1708.06047}{{\ttfamily
  1708.06047}}].

\bibitem{Witten:1979ey}
E.~Witten, \emph{{Dyons of Charge e theta/2 pi}},
  \href{https://doi.org/10.1016/0370-2693(79)90838-4}{\emph{Phys. Lett.}
  {\bfseries B86} (1979) 283}.

\bibitem{Fischler:1983sc}
W.~Fischler and J.~Preskill, \emph{{DYON - AXION DYNAMICS}},
  \href{https://doi.org/10.1016/0370-2693(83)91260-1}{\emph{Phys. Lett.}
  {\bfseries 125B} (1983) 165}.

\bibitem{Lazarides:1982tw}
G.~Lazarides and Q.~Shafi, \emph{{Axion Models with No Domain Wall Problem}},
  \href{https://doi.org/10.1016/0370-2693(82)90506-8}{\emph{Phys. Lett.}
  {\bfseries 115B} (1982) 21}.

\bibitem{Leonhardt:2000km}
U.~Leonhardt and G.~E. Volovik, \emph{{How to create Alice string (half quantum
  vortex) in a vector Bose-Einstein condensate}},
  \href{https://doi.org/10.1134/1.1312008}{\emph{Pisma Zh. Eksp. Teor. Fiz.}
  {\bfseries 72} (2000) 66}
  [\href{https://arxiv.org/abs/cond-mat/0003428}{{\ttfamily
  cond-mat/0003428}}].

\bibitem{Chatterjee:2017jsi}
C.~Chatterjee and M.~Nitta, \emph{{BPS Alice strings}},
  \href{https://doi.org/10.1007/JHEP09(2017)046}{\emph{JHEP} {\bfseries 09}
  (2017) 046} [\href{https://arxiv.org/abs/1703.08971}{{\ttfamily
  1703.08971}}].

\bibitem{Kiskis:1978ed}
J.~E. Kiskis, \emph{{Disconnected Gauge Groups and the Global Violation of
  Charge Conservation}},
  \href{https://doi.org/10.1103/PhysRevD.17.3196}{\emph{Phys. Rev.} {\bfseries
  D17} (1978) 3196}.

\bibitem{Bucher:1992bd}
M.~Bucher, H.-K. Lo and J.~Preskill, \emph{{Topological approach to Alice
  electrodynamics}},
  \href{https://doi.org/10.1016/0550-3213(92)90173-9}{\emph{Nucl. Phys.}
  {\bfseries B386} (1992) 3}
  [\href{https://arxiv.org/abs/hep-th/9112039}{{\ttfamily hep-th/9112039}}].

\bibitem{Schwarz:1982ec}
A.~S. Schwarz, \emph{{FIELD THEORIES WITH NO LOCAL CONSERVATION OF THE ELECTRIC
  CHARGE}}, \href{https://doi.org/10.1016/0550-3213(82)90190-0}{\emph{Nucl.
  Phys.} {\bfseries B208} (1982) 141}.

\bibitem{Blinnikov:1982eh}
S.~I. Blinnikov and M.~{\relax Yu}. Khlopov, \emph{{ON POSSIBLE EFFECTS OF
  'MIRROR' PARTICLES}}, {\emph{Sov. J. Nucl. Phys.} {\bfseries 36} (1982) 472}.

\bibitem{Kim:1979if}
J.~E. Kim, \emph{{Weak Interaction Singlet and Strong CP Invariance}},
  \href{https://doi.org/10.1103/PhysRevLett.43.103}{\emph{Phys. Rev. Lett.}
  {\bfseries 43} (1979) 103}.

\bibitem{Shifman:1979if}
M.~A. Shifman, A.~I. Vainshtein and V.~I. Zakharov, \emph{{Can Confinement
  Ensure Natural CP Invariance of Strong Interactions?}},
  \href{https://doi.org/10.1016/0550-3213(80)90209-6}{\emph{Nucl. Phys.}
  {\bfseries B166} (1980) 493}.

\bibitem{Dine:1981rt}
M.~Dine, W.~Fischler and M.~Srednicki, \emph{{A Simple Solution to the Strong
  CP Problem with a Harmless Axion}},
  \href{https://doi.org/10.1016/0370-2693(81)90590-6}{\emph{Phys. Lett.}
  {\bfseries 104B} (1981) 199}.

\bibitem{Zhitnitsky:1980tq}
A.~R. Zhitnitsky, \emph{{On Possible Suppression of the Axion Hadron
  Interactions. (In Russian)}}, {\emph{Sov. J. Nucl. Phys.} {\bfseries 31}
  (1980) 260}.

\bibitem{Kawasaki:2015ofa}
M.~Kawasaki, M.~Yamada and T.~T. Yanagida, \emph{{Observable dark radiation
  from a cosmologically safe QCD axion}},
  \href{https://doi.org/10.1103/PhysRevD.91.125018}{\emph{Phys. Rev.}
  {\bfseries D91} (2015) 125018}
  [\href{https://arxiv.org/abs/1504.04126}{{\ttfamily 1504.04126}}].

\bibitem{Kibble:1976sj}
T.~W.~B. Kibble, \emph{{Topology of Cosmic Domains and Strings}},
  \href{https://doi.org/10.1088/0305-4470/9/8/029}{\emph{J. Phys.} {\bfseries
  A9} (1976) 1387}.

\bibitem{Cline:2013gha}
J.~M. Cline, K.~Kainulainen, P.~Scott and C.~Weniger, \emph{{Update on scalar
  singlet dark matter}}, \href{https://doi.org/10.1103/PhysRevD.92.039906,
  10.1103/PhysRevD.88.055025}{\emph{Phys. Rev.} {\bfseries D88} (2013) 055025}
  [\href{https://arxiv.org/abs/1306.4710}{{\ttfamily 1306.4710}}].

\bibitem{Bruemmer:2009ky}
F.~Brummer, J.~Jaeckel and V.~V. Khoze, \emph{{Magnetic Mixing: Electric
  Minicharges from Magnetic Monopoles}},
  \href{https://doi.org/10.1088/1126-6708/2009/06/037}{\emph{JHEP} {\bfseries
  06} (2009) 037} [\href{https://arxiv.org/abs/0905.0633}{{\ttfamily
  0905.0633}}].

\bibitem{DelNobile:2015bqo}
E.~Del~Nobile, M.~Nardecchia and P.~Panci, \emph{{Millicharge or Decay: A
  Critical Take on Minimal Dark Matter}},
  \href{https://doi.org/10.1088/1475-7516/2016/04/048}{\emph{JCAP} {\bfseries
  1604} (2016) 048} [\href{https://arxiv.org/abs/1512.05353}{{\ttfamily
  1512.05353}}].

\bibitem{Akerib:2016vxi}
{\scshape LUX} collaboration, D.~S. Akerib et~al., \emph{{Results from a search
  for dark matter in the complete LUX exposure}},
  \href{https://doi.org/10.1103/PhysRevLett.118.021303}{\emph{Phys. Rev. Lett.}
  {\bfseries 118} (2017) 021303}
  [\href{https://arxiv.org/abs/1608.07648}{{\ttfamily 1608.07648}}].

\bibitem{Turner:1986tb}
M.~S. Turner, \emph{{Thermal Production of Not SO Invisible Axions in the Early
  Universe}}, \href{https://doi.org/10.1103/PhysRevLett.59.2489,
  10.1103/PhysRevLett.60.1101.3}{\emph{Phys. Rev. Lett.} {\bfseries 59} (1987)
  2489}.

\bibitem{Masso:2002np}
E.~Masso, F.~Rota and G.~Zsembinszki, \emph{{On axion thermalization in the
  early universe}},
  \href{https://doi.org/10.1103/PhysRevD.66.023004}{\emph{Phys. Rev.}
  {\bfseries D66} (2002) 023004}
  [\href{https://arxiv.org/abs/hep-ph/0203221}{{\ttfamily hep-ph/0203221}}].

\bibitem{Ade:2015xua}
{\scshape Planck} collaboration, P.~A.~R. Ade et~al., \emph{{Planck 2015
  results. XIII. Cosmological parameters}},
  \href{https://doi.org/10.1051/0004-6361/201525830}{\emph{Astron. Astrophys.}
  {\bfseries 594} (2016) A13}
  [\href{https://arxiv.org/abs/1502.01589}{{\ttfamily 1502.01589}}].

\bibitem{Wu:2014hta}
W.~L.~K. Wu, J.~Errard, C.~Dvorkin, C.~L. Kuo, A.~T. Lee, P.~McDonald et~al.,
  \emph{{A Guide to Designing Future Ground-based Cosmic Microwave Background
  Experiments}},
  \href{https://doi.org/10.1088/0004-637X/788/2/138}{\emph{Astrophys. J.}
  {\bfseries 788} (2014) 138}
  [\href{https://arxiv.org/abs/1402.4108}{{\ttfamily 1402.4108}}].

\bibitem{Abazajian:2013oma}
{\scshape Topical Conveners: K.N. Abazajian, J.E. Carlstrom, A.T. Lee}
  collaboration, K.~N. Abazajian et~al., \emph{{Neutrino Physics from the
  Cosmic Microwave Background and Large Scale Structure}},
  \href{https://doi.org/10.1016/j.astropartphys.2014.05.014}{\emph{Astropart.
  Phys.} {\bfseries 63} (2015) 66}
  [\href{https://arxiv.org/abs/1309.5383}{{\ttfamily 1309.5383}}].

\bibitem{Huang:2018fum}
D.~Huang and B.-Q. Lu, \emph{{Comment on “Hearing the signal of dark sectors
  with gravitational wave detectors”}},
  \href{https://doi.org/10.1103/PhysRevD.97.068303}{\emph{Phys. Rev.}
  {\bfseries D97} (2018) 068303}
  [\href{https://arxiv.org/abs/1803.03180}{{\ttfamily 1803.03180}}].

\end{thebibliography}\endgroup

\end{document}